# Bi-Objective Community Detection (BOCD) in Networks using Genetic Algorithm

Rohan Agrawal

Jaypee Institute of Information Technology, Computer Science Department,
Noida - 201307, Uttar Pradesh, India
rohan.agrawal.89@jiitu.org

**Abstract.** A lot of research effort has been put into community detection from all corners of academic interest such as physics, mathematics and computer science. In this paper I have proposed a Bi-Objective Genetic Algorithm for community detection which maximizes modularity and community score. Then the results obtained for both benchmark and real life data sets are compared with other algorithms using the modularity and MNI performance metrics. The results show that the BOCD algorithm is capable of successfully detecting community structure in both real life and synthetic datasets, as well as improving upon the performance of previous techniques.

## 1   Introduction

In the context of networks, community structure refers to the occurrence of groups of nodes in a network that are more densely connected than with the rest of the nodes in the network. The inhomogeneous connections suggest that the network has certain natural division within it.

The occurrence of community structure is quite common in real networks. An example of the occurrence of community structure in real networks is the appearance of groups in social networks. Let's take the example of a social networking site. Let a node represent an individual and let the edge represent friendship relation between two individuals. If many students in a particular class or school are friends among themselves, then the network graph will have many connections between them. Thus one community could be identified as a school community. Other communities could be related to work, family, colleges or common interests.

Other examples are citation networks which form communities by research topics. Sport teams form communities on the basis of the division in which they play, as they will play more often with teams that are in the same division/community as them.
Now let us consider the potential applications of the detection of communities in networks. Communities in a social network might help us find real social groupings, perhaps by interest or background. Communities can have concrete applications. Clustering Web clients who have similar interests and are geographically near to each other may improve the performance of services provided on the World Wide Web, in that each cluster of clients could be served by a dedicated mirror server [1]. Identifying clusters of customers with similar interests in the network of purchase relationships between customers and products of online retailers enables to set up efficient recommendation systems [2], that better guide customers through the list of items of the retailer and enhance the business opportunities. Clusters of large graphs can be used to create data structures in order to efficiently store the graph data and to handle navigational queries, like path searches [3][4]. Ad hoc networks [5], i.e. self-configuring networks formed by communication nodes acting in the same region and rapidly changing (because the devices move, for instance), usually have no centrally maintained routing tables that specify how nodes have to communicate to other nodes. Grouping the nodes into clusters enables one to generate compact routing tables while the choice of the communication paths is still efficient [6].

The aim of community detection in graphs is to identify the modules by using the information encoded in the network topology. Weiss and Jacobson [7] were among the first to analyze community structure. They searched for work groups within a government agency. Already in 1927, Stuart Rice looked for clusters of people in small political bodies based on the similarity of their voting patterns [8].

In a paper appearing in 2002, Girvan and Newman proposed a new algorithm, aiming at the identification of edges lying between communities and their successive removal. After a few iterations, this process led to the isolation of communities [9]. The paper triggered inertest in this field, and many new methods have been proposed in previous years.

In particular, physicists entered the game, bringing in their tools and techniques: spin models, optimization, percolation, random walks, synchronization, etc., became ingredients of new original algorithms. The field has also taken advantage of concepts and methods from computer science, nonlinear dynamics, sociology, discrete mathematics.

Genetic algorithms [10] have also been used to optimize modularity. In a standard genetic algorithm one has a set of candidate solutions to a problem, which are numerically encoded as chromosomes, and an objective function to be optimized on the space of solutions. The objective function plays the role of biological fitness for the chromosomes. One usually starts from a random set of candidate solutions, which are progressively changed through manipulations inspired by biological processes regarding real chromosomes, like point mutation (random variations of some parts of the chromosome) and crossing over (generating new chromosomes by merging parts of existing chromosomes). Then, the fitness of the new pool of candidates is computed and the chromosomes with the highest fitness have the greatest chances to survive in the next generation. After several iterations only solutions with large fitness survive. In a work by Tasgin et al. [11], partitions are the chromosomes and modularity is the fitness function.

Genetic algorithms were also adopted by Liu et al. [12]. Here the maximum modularity partition is obtained via successive bipartitions of the graph, where each bipartition is determined by applying a genetic algorithm to each sub graph (starting from the original graph itself), which is considered isolated from the rest of the graph. A bipartition is accepted only if it increases the total modularity of the graph.

In 2009, Pizutti [13] proposed a multi-objective genetic algorithm for the detection of communities in a network. The two fitness functions used were community score and community fitness. The algorithm had the advantage that it provided a set of solutions based on the maximization of both the evaluation functions.

In section 2, the problem of Community Detection will be formulated mathematically with the introduction of two functions. In section 3, all the stages of the Genetic Algorithm such as Initialization, Fitness Functions, Mutation and Crossover will be elaborated upon. In section 4, the experimental results of BOCD will be presented and compared with existing Community Detection techniques. The Conclusion will be presented in section 5.

## 2 Problem Definition

A network $N_w$ can be modeled as a graph $G = (V,E)$ where $V$ is a set of objects, called nodes or vertices, and $E$ is a set of links, called edges, that connect two elements of $V$. A community (or cluster) in a network is a group of vertices having a high density of edges within them, and a lower density of edges between groups. The problem of detecting $k$ communities in a network, where the number $k$ is unknown, can be formulated as finding a partitioning of the nodes in $k$ subsets that are highly intra-connected and sparsely inter-connected. To deal with graphs, often the adjacency matrix is used. If the network is constituted by $N$ nodes, the graph can be represented with the $N \times N$ adjacency matrix $A$, where the entry at position $(i, j)$ is 1 if there is an edge from node $i$ to node $j$, 0 otherwise.

Let us introduce the concept of Community Score as a defined in [13] and [14]. Let $S \subset G$ be the sub graph where node $i$ belongs to, the degree of $i$ with respect to $S$ can be split as

$$k_i(S) = k_{in}^i(S) + k_{out}^i(S) .$$

where

$$k_{in}^i(S) = \sum_{j \in S} A_{ij} .$$

is the number of edges connecting i to the other nodes in $S$. Here $A$ is the adjacency matrix of $G$.

$$k_{out}^i(S) = \sum_{j \notin S} A_{ij} .$$

is the number of edges connecting i to the rest of the network. Let $\mu_i$ represent the fraction of edges connecting $i$ to the other nodes in $S$.

$$\mu_i = \frac{1}{|S|} k_i^{in}(S).$$

where $|S|$ is the cardinality of $S$. The power mean of $S$ of order $r$, $M(s)$

$$M(s) = \frac{\sum_{i \in S}(\mu_i)^r}{|S|}.$$

In the computation of M(s), since $0 \leq \mu_i \leq 1$, the exponent r increases the weight of nodes having many connections with other nodes belonging to the same community, and diminishes the weight of those nodes having few connections inside S.

The volume $v_s$ of a community is defined as the number of edges connecting vertices inside S,

$$v_s = \sum_{i,j \in S} A_{ij}.$$

The *score* of $S$ is defined as

$$score(S) = M(S) \times v_s.$$

The Community score of a clustering $\{S_1, \ldots S_k\}$ of a network is defined as

$$CS = \sum_{i=1}^{k} score(S_i). \qquad (1)$$

The problem of community detection has been formulated in [14] as the problem of maximizing the Community Score. The other objective is to maximize modularity, defined in [15]. Let k be the number of modules found inside a network. The modularity is defined as

$$Q = \sum_{s=1}^{k} \left[ \frac{l_s}{m} - \left(\frac{d_s}{2m}\right)^2 \right]. \qquad (2)$$

where $l_s$ is the total number of edges joining vertices inside the module s, and $\frac{l_s}{m}$ represents the fraction of edges in the network that connect the same community. $d_s$ represents the sum of the degrees of the nodes of s. If the number of within-community edges is no more than random, we will get $Q = 0$. The maximum value of Q is 1, which indicates strong community structure.

## 3 Algorithm Description

The various stages of the genetic algorithm have been described in the following subsections. The framework used was NSGA-II in C described in [24].

### 3.1 Genetic Representation

The chromosome is represented in the format mentioned in [16]. The representation of an individual consists of N genes, and each gene can take a value in the range {1, ..., N}, where N is the number of nodes in the network. If a value j is assigned to the $i^{th}$ gene, this suggests that i and j are in the same cluster. But if i and j are already assigned, then the gene is ignored. Thus later genes will have less bearing on cluster formation. The decoding of this individual to obtain clusters can be done in linear time according to [17].

For example, consider the individual for a network of 34 nodes (N = 34). The number in the curly brackets represents the index of the element in the individual.

{1}2, {2}3, {3}4, {4}14, {5}17, {6}17, {7}6, {8}14, {9}19, {10}19, {11}17, {12}14, {13}2, {14}9, {15}19, {16}9, {17}15, {18}8, {19}21, {20}8, {21}27, {22}1, {23}15, {24}26, {25}26, {26}32, {27}30, {28}26, {29}25, {30}3, {31}19, {32}4, {33}23, {34}9

We are assuming here that if the above individual was stored in an array, the index of the 1st element would be 1 and not 0. In the above chromosome, the element at index 1 of the array is 2. Thus nodes 1 and 2 are in the same cluster. Similarly the element at index position 2 is 3, thus 2 and 3 are put in the same cluster. Since nodes 1 and 2 are already in Cluster 1, we have nodes 1, 2 and 3 put in the same cluster. The element at index position 3 is 4, thus nodes 3 and 4 are also in the same cluster. The element at the 5th index position is 17. Since neither 5, nor 17 have been previously assigned a cluster, they are put together in a new cluster, Cluster 2. This process goes on iteratively till the last element. Finally the clusters are made as follows:

Cluster1:   1, 2, 3, 4, 14, 8, 12, 13, 18, 20, 22
Cluster2:   5, 17, 6, 7, 11
Cluster3:   9, 19, 10, 15, 16, 21, 27, 23, 30, 31, 33, 34
Cluster4:   24, 26, 25, 32, 28, 29

### 3.2 Initialization

The population is initialized randomly from values between 1 and N, where N is the number of nodes in the network.

### 3.3 Fitness Functions

The algorithm used here is a bi-objective optimization, where both fitness functions are minimized. The first fitness function is derived from equation (2).

$$\min f_1 = 1 - Q.$$

The 2nd fitness function uses both equations (1) and (2).

$$\min f_2 = (1 - Q) + (\frac{10}{1 + CS})$$

$Q$ lies in the range [0,1], therefore the minimization of $(1 - Q)$ helps in finding the maximum value of modularity. In the second fitness function, the weight 10 for the Community Structure term ($CS$) has been found out empirically. The above pair of fitness functions taken together performs better than the single objective optimization of either of the two taken separately.

### 3.4 Crossover and Mutation

Simple Uniform crossover is used as the crossover operator. The crossover site is chosen at random. Selection strategy used is tournament selection, with 4 individuals contesting in the tournament.

Take for e.g.
Parent 1:    1, 2, 4, 5, 3, 5, 6, 1, 9, 4
Parent 2:    3, 6, 3, 2, 6, 4, 3, 1, 2, 9

Suppose the crossover site is randomly decided at 5. This means that the first 5 elements of Child 1 will come from Parent 1, i.e. {1, 2, 4, 5, 3}. The other elements for Child 1 will come from Parent 2, i.e. {4, 3, 1, 2, 9}. The beginning elements for Child 2 come from Parent 2 and the latter elements come from Parent 1. Thus the children formed are:

Child 1:    1, 2, 4, 5, 3, 4, 3, 1, 2, 9
Child 2:    3, 6, 3, 2, 6, 5, 6, 1, 9, 4

Mutation operator also performs simple mutation, i.e. a gene is chosen at random and its value is simply changed.

## 4 Experimental Results

Bi-Objective Community Detection (BOCD) is applied on 3 real world networks, the American College Football [19], Bottlenose Dolphin [26] and the Zachary Karate Club [18] network. The method is also tested on a benchmark generating program proposed in [23] which is an extension of the benchmark proposed by Girvan and Newman in [9].

The experiments were performed on a Core2duo machine, 2.0 Giga Hz with 3 Mb RAM. The framework for Multi-Objective Genetic Algorithm used was NSGA-II written in C described in [24]. The parameters used in compiling the code are as follows:

Population: 200
Generations: 3000
Crossover Probability: 0.7
Mutation Probability: 0.03

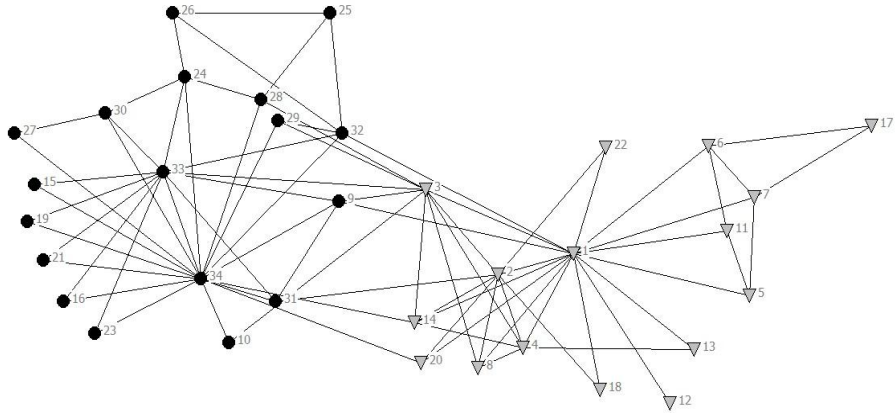

**Fig. 1.** The 34 node Zachary Karate Club Network divided into 2 communities. This was how the club actually broke into 2 groups. The first group is shown by circular nodes and the second by triangular nodes. The modularity of this division is 0.371.

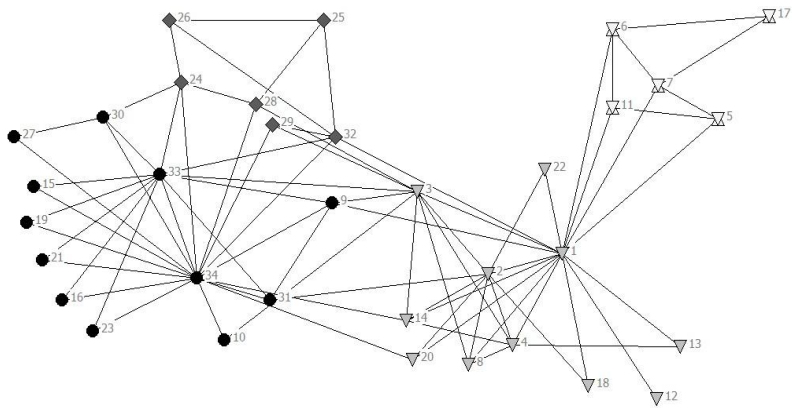

**Fig. 2.** Division of the Zachary Karate Club Network into 4 communities by BOCD. Each community is shown with a different symbol. The modularity of this division is 0.419.

The evaluation metrics used were Modularity which was described above, as well as Normalized Mutual Information which was described in [22]. The results obtained by BOCD are compared with the fast GN algorithm [20] and MOGA-Net [13] on the basis of Modularity and NMI.

**4.1 Zachary Karate Club Network**

This network was generated by Zachary [18], who studied the friendship of 34 members of a karate club over a period of two years. During this period, because of disagreements, the club divided in two groups almost of the same size. The original division of the club in 2 communities is shown in Figure 1. The BOCD algorithm divides the nodes into 4 communities, with this separation showing a higher value of modularity then the original solution itself. As can be seen from Table 1, BOCD performs better than both GN and MOGA-Net in terms of modularity. The NMI of the division was found to be 0.695622, which is better than GN algorithm but not MAGA-Net. As MOGA-Net generates a pareto set of results, they have achieved higher NMI values.

### 4.2 American College Football Network

The American College Football network [9] is a network of 115 teams, where the edges represent the regular season games between the two teams they connect. The teams are divided into conferences and play teams within their own conference more frequently. The network has 12 conferences or communities. The division obtained by BOCD was better than the result of MOGA-Net and was exactly on equal terms with the modularity value of the GN algorithm. The NMI of the division was found to be 0.878178, which is the highest value among the three algorithms.

### 4.3 Bottlenose Dolphin Network

The network of 62 bottlenose dolphins living in Doubtful Sound, New Zealand, was compiled in [26] by Lusseau from seven years of dolphin behavior. A tie between 2 dolphins was established by their statistically frequent association. The network split naturally into 2 large groups, the number of ties being 159. The performance of BOCD was much better than the GN algorithm and marginally better than that of MOGA-Net. The NMI of the division was found to be 0.615492, which lies in between MOGA-Net and GN performance wise, the best being MOGA-Net.

**Table 1.** Comparison of Modularity values for the 3 real datasets. The first column gives the value of modularity for NMI = 1. The following columns give modularity results for the fast GN, MOGA-Net and BOCD algorithms.

| Dataset | Mod. For NMI=1 | GN | MOGA | BOCD |
|---|---|---|---|---|
| Zachary Karate Club | 0.371 | 0.380 | 0.415 | 0.419 |
| College Football | 0.518 | 0.577 | 0.515 | 0.577 |
| Bottlenose Dolphins | 0.373 | 0.495 | 0.505 | 0.507 |

**Table 2.** Comparison of NMI values for the 3 real datasets. The columns give NMI results for the fast GN, MOGA-Net and BOCD algorithms.

| Dataset | GN | MOGA | BOCD |
|---|---|---|---|
| Zachary Karate Club | 0.692 | 1.0 | 0.695 |
| College Football | 0.762 | 0.795 | 0.878 |
| Bottlenose Dolphins | 0.573 | 1.0 | 0.615 |

## 4.4 Benchmark Test Network

The network consists of 128 nodes divided into four communities of 32 nodes each. The average degree of each node is 16. The fraction of edges shared by each node with nodes in its own community is known as the mixing parameter. If the value of the mixing parameter $\mu > 0.5$, it suggests that a node will have more link to other nodes, outside its community. Thus finding community structure will be difficult for $\mu = 0.5$, as evident from the following graph. According to a graph drawn in [13], MOGA-Net could achieve an NMI of less than 0.1 for $\mu = 0.5$. Thus our algorithm performs better in case of a higher mixing parameter.

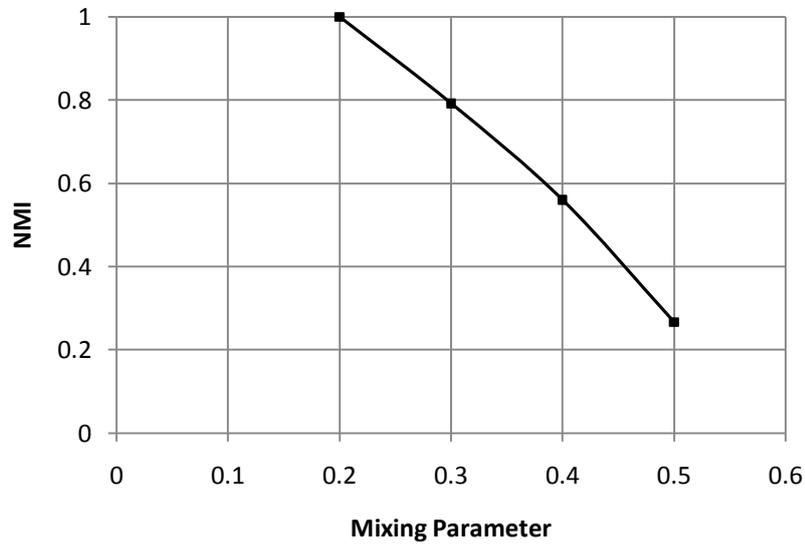

**Fig. 3.** NMI values obtained by BOCD for different values of mixing parameter. Here r=2.5.

**Table 3.** Modularity and NMI values for benchmark network with increasing value of Mixing Parameter. The performance for $\mu = 0.2$ is the best, and expectedly deteriorates as $\mu$ is increased.

| Mixing Parameter($\mu$) | Modularity | NMI |
|---|---|---|
| 0.2 | 0.4511 | 1.0 |
| 0.3 | 0.347 | 0.792138 |
| 0.4 | 0.218 | 0.559844 |
| 0.5 | 0.181 | 0.266481 |

## 5   Conclusions

The paper presented a Bi-Objective Community Detection technique through the use of Genetic Algorithm. By simply combining community score and modularity, the BOCD algorithm improved upon the performance of both the GN algorithm which used Modularity and MOGA-Net which used community score in the community detection problem. Results on real life networks as well as synthetic benchmarks show the capability of this approach in finding out communities within networks. Future research should aim at decreasing computational complexity of Community Detecting algorithms and finding communities in networks with a high mixing parameter.

**Acknowledgments.** I would like to thank Dr S K. Gupta and Pranava Chaudhary, both of whom have profoundly helped me in understanding Genetic Algorithm and Multi-Objective Optimization. I would also like to thank Andrea Lancichinetti for providing the NMI calculating code. The software used for drawing network diagrams was Netdraw [25].